\documentclass{article}

    \PassOptionsToPackage{numbers, compress}{natbib}
\usepackage[final]{neurips_2020_ml4ps}
    
\usepackage[utf8]{inputenc} % allow utf-8 input
\usepackage[T1]{fontenc}    % use 8-bit T1 fonts
\usepackage{hyperref}       % hyperlinks
\usepackage{url}            % simple URL typesetting
\usepackage{booktabs}       % professional-quality tables
\usepackage{amsfonts}       % blackboard math symbols
\usepackage{nicefrac}       % compact symbols for 1/2, etc.
\usepackage{microtype}      % microtypography
\usepackage{graphicx}
\usepackage{braket}
\usepackage{amsmath}

\DeclareMathOperator*{\argmin}{arg\,min}

\usepackage{xcolor}

\title{Scalable variational Monte Carlo \\
with graph neural ansatz}

\author{%
  Li Yang \\
  Google Research \\
  Mountain View, CA 94043, USA \\
  \texttt{lyliyang@google.com} \\
  \And
  Wenjun Hu \\
  Department of Physics and Astronomy \\ University of Tennessee \\ 
  Knoxville, TN 37996, USA \\
  \texttt{whu11@utk.edu} \\
  \And
  Li Li \\
  Google Research \\
  Mountain View, CA 94043, USA \\
  \texttt{leeley@google.com} \\
}

\begin{document}

\maketitle

\begin{abstract}
Deep neural networks have been shown as a potentially powerful ansatz in variational Monte Carlo for solving quantum many-body problems. We propose two improvements in this direction. The first is graph neural ansatz (GNA), which is a variational wavefunction universal to arbitrary geometry. GNA results in accurate ground-state energies on 2D Kagome lattices, triangular lattices, and randomly connected graphs.
Secondly, we design a distributed workflow on multiple accelerators to scale up the computation. We compute Kagome lattices with sizes up to 432 sites on 128 TPU cores. The parameter sharing nature of the GNA also leads to transferability across different system sizes and geometries.
\end{abstract}

\section{Introduction}\label{intro}

Quantum many-body systems have been studied for a century but still with many unsolved problems. Many numerical algorithms have been developed. For example, the tensor network~\cite{white1992density,SCHOLLWOCK201196, ORUS2014117} based methods have been successfully used for many low-dimensional systems, but their applications in two-dimensional and above are still under development. The quantum Monte Carlo (QMC) methods based on probabilistic sampling typically require a positive-semidefinite wavefunction~\cite{CEPERLEY:1986aa, Foulkes:2001aa}. Variational Monte Carlo (VMC)~\cite{McMillan:1965aa} works well for arbitrary dimension and wavefunction signs.
It requires a variational ansatz for the many-body wavefunction, which is a multi-dimensional complex-valued function whose input is a many-body state configuration and output is its amplitude.
Previously, physicists designed them with the physical insight of the system~\cite{becca2017quantum}, and in a form with much fewer parameters comparing with today's deep neural networks (DNN).
Pioneer work by \citet{Carleo:2017aa} proposed to use a restricted Boltzmann machine (RBM) as the variational ansatz. Following this direction, the RBM and a few other shallow networks have been applied to study several quantum many-body systems with good accuracy~\cite{Nomura:2017aa, Saito_2018, Cai:2018aa, Liang:2018aa, Choo:2018aa, Kaubruegger:2018aa}. Recently, DNNs has been used as variational ansatzes~\cite{kochkov2018variational, Sharir:2020aa, Choo:2019aa, yang2020deep} and optimized with algorithms tailored to deep learning.

In this work, we incorporate two recent advances in deep learning into VMC, making it universal to the geometry of the system and scalable in system size. The first component is a graph neural ansatz (GNA) wavefunction, i.e. a graph neural network (GNN)~\cite{battaglia2018Relational} based variational ansatz. It can encode arbitrary geometries directly, while previously used convolutional neural networks (CNN) require additional artifacts to fit non-square lattice into grids. GNNs have been receiving increasing attention and have been successfully applied to many areas including social networks~\cite{NIPS2017_6703, kipf2016semi}, computer vision~\cite{7974879}, combinatorial optimization~\cite{NIPS2017_7214}, and more recently applications in physical sciences such as quantum chemistry~\cite{gilmer2017neural,kearnes2019decoding,hu2019strategies}, classical physics engines~\cite{pmlr-v80-sanchez-gonzalez18a,Battaglia2016InteractionNF,sanchez2020learning}, glassy systems~\cite{bapst2020unveiling} and protein interactions ~\cite{NIPS2017_7231}. As a demonstration of GNA, we use graph convolutional networks (GCN)~\cite{kipf2016semi} for various kinds of geometries including the 2D Kagome lattices, triangular lattices, and randomly connected graphs.
The second component is a scalable implementation of importance sampling gradient optimization (ISGO)~\cite{yang2020deep} on multiple accelerators (GPUs, TPUs). This makes the computation easily scaling up to $12\times12\times3$ Kagome lattices, which is typically impossible to fit into a single accelerator device.

Major contributions: 1. Proposed GNA as a universal ansatz wavefunction for VMC. 2. A design of the scalable implementation of VMC on accelerators, demonstrated on 2D Kagome lattice up to 432 sites with 128 TPU cores. 3. Showed transferability of GNA across different system sizes and geometries.

\begin{figure*}[t!]
	\includegraphics[width=0.95\textwidth]{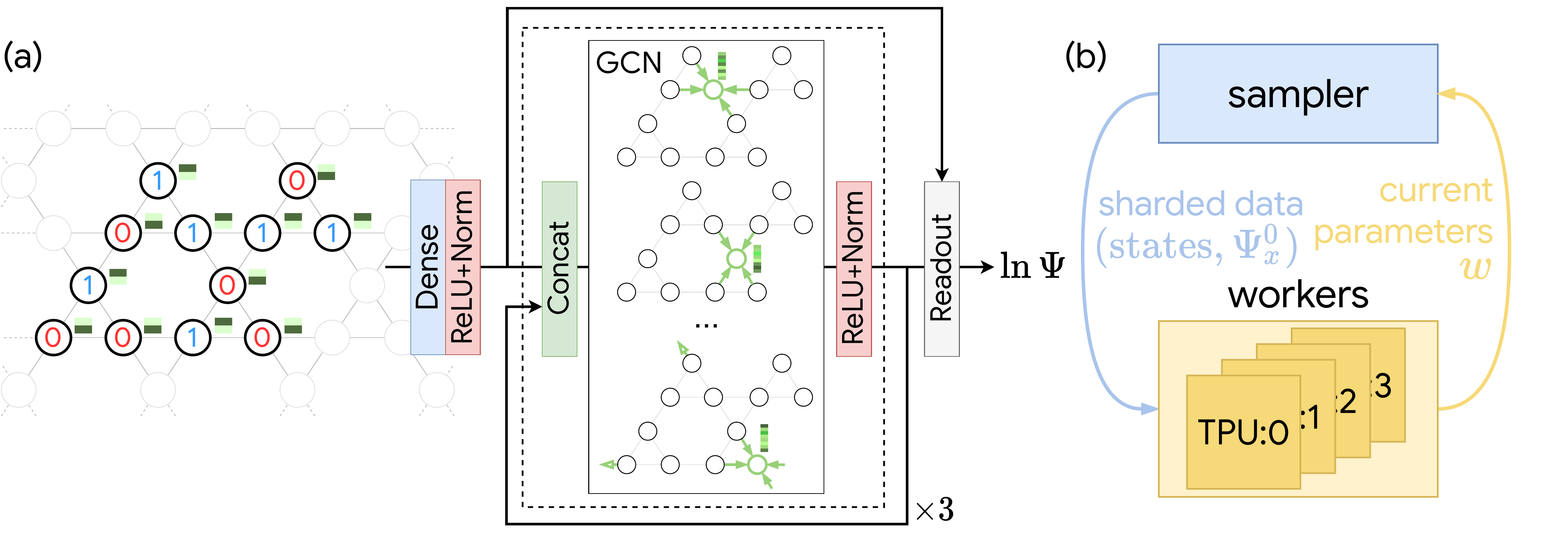}
	\centering
	\caption{(a) Illustration of graph neural ansatz. (b) Distributed workflow of VMC using ISGO. }
	\label{fig:vmc_graph}
\end{figure*}

\section{Graph neural ansatz for quantum many-body system}\label{gcn}

Existing studies of quantum many-body systems using VMC usually constrain on systems with structured geometries, e.g. with translational or rotational symmetries. Certain symmetry consideration is essential to the novel property of the system and brings intuition on physicists' design of ansatz. Recently, there is a trend to use DNNs as variational ansatzes since neural networks are universal approximators. Most of the applications are on 1D chains or 2D square lattices using RBMs or CNNs. Libraries such as NetKet~\cite{CARLEO2019100311} support non-square geometries, but do not generalize to arbitrary geometries nor run on accelerators. CNN has been shown to work well for approximating many-body ground states~\cite{yang2020deep, Liang:2018aa, Choo:2019aa}. We attribute the success of CNNs to its inductive bias of capturing local connections and sharing of parameters. On the other hand, GNN can contain the inductive bias of capturing arbitrary connections, and also sharing of parameters among nodes and edges. In this paper, we use a special kind of GNN, GCN~\cite{kipf2016semi}, as the GNA for various system geometries. In a GCN, each node has an embedding vector.
Node embeddings at layer $l$ are updated by 
\begin{equation}\label{eq:gcn}
    H^{(l+1)} = f([D^{-\frac{1}{2}} A D^{-\frac{1}{2}} H^{(l)}; H^{(l)}]W + b),
\end{equation}
where $H^{(l)}\in\mathbb{R}^{N_\mathrm{site}\times F}$ is the node embeddings at layer $l$, $N_\mathrm{site}$ the total number of sites in the system, $F$ the embedding size, $A$ the adjacency matrix and $D$ is a node degree diagonal matrix for normalizing the adjacency matrix. 
We also concatenate the messages from neighboring nodes with current embeddings along feature dimension and then linearly transform using $W\in\mathbb{R}^{2F\times F}$ and $b\in\mathbb{R}^{F}$. Note $W$ and $b$ are shared across GCN layers. Finally we apply nonlinear transformation $f(\cdot)$ using ReLU~\cite{nair2010rectified} followed by a layer normalization~\cite{ba2016layer} to get the node embeddings at layer $l + 1$.

Fig~\ref{fig:vmc_graph}(a) illustrates GNA on a Kagome lattice of size $2\times2\times3$. The periodic boundary conditions are enforced while constructing $A$. The sites on one edge of the system connect to the sites on the opposite edge.
The state $x$, occupation of Bosons on each site $\{0,1\}$, is one-hot encoded and then nonlinearly transformed to embeddings with size $F=16$. Graph convolution in Eq.~(\ref{eq:gcn}) is recurrently applied three times. Finally, we concatenate the output with input, sum up the embeddings overall sites to a single embedding vector, and map to a scalar $\ln\Psi$ using a dense layer of one unit.

\section{Scaling up VMC to multiple accelerators}\label{scale-up}

Consider a neural network $\Psi(x;w)$ as the variational ansatz with a set of trainable parameters $w$.
The ansatz with optimal parameters $w^* = \argmin_w E_v(w)$ approximates the ground state wavefunction, where the $E_v(w)=\bra{\Psi(x;w)} H \ket{\Psi(x;w)} / \braket{\Psi(x;w) | \Psi(x;w)}$ is the variational energy.
However, direct computation of $E_v(w)$ is infeasible due to the exponential size of the Hilbert space.
A stochastic optimization approach is used. First $N_{\mathrm{sample}}$ quantum states following distribution $P^0_x \propto |\Psi^0_x|^2$ are sampled using the Markov-chain Monte Carlo (MCMC) method. Here $\Psi^0$ is the wavefunction under the current parameter $w$, and $x$ indexes the sampled states. Then the gradients $\partial_w E_v$ can be approximated stochastically,
\begin{equation}
\label{vmc}
\begin{split}
\partial_{w}E_v \approx \sum_x I^0 E^0_x\partial_w\text{ln}\Psi^0_x
- \sum_x I^0 E^0_x\sum_x I^0 \partial_w\text{ln}\Psi^0_x,
\end{split}
\end{equation}
where $E^0_x = \sum_{x'}H_{x,x'}\Psi^0_{x'}/\Psi^0_x$ is defined as the local energy under $\Psi^0$ and $I^0 = 2 / N_\mathrm{sample}$. Then the parameters are updated by $w \leftarrow w - \alpha \partial_w E_v$. In conventional VMC, after the parameters are updated once, another iteration of sampling begins. This process makes VMC highly inefficient while using DNNs as the variational ansatz. Since the MCMC sampling must be long enough to obtain a faithful sample distribution following $P^0_x$. The ISGO method~\cite{yang2020deep} has been developed to overcome this issue by reusing samples for parameters updating. Note that it is incorrect to simply use the same set of samples to perform multiple updates with Eq.~(\ref{vmc}), since after each update, the samples following distribution of old wavefunction will not follow the distribution of the new wavefunction. The key to resolve this is to renormalize the distribution of those mismatched samples to $P^0_x$ by multiplying the local energies and derivatives in Eq.~(\ref{vmc}) with importance sampling:
\begin{equation} 
\label{isgo}
\begin{split}
\partial_{w}E_v \approx \sum_x I_x E_x\partial_w\text{ln}\Psi_x
- \sum_x I_x E_x\sum_x I_x \partial_w\text{ln}\Psi_x,
\end{split}
\end{equation}
where $E_x$ is the local energy under the new wavefunction $\Psi$, and $I_x/I^0 = P_x/P^0_x = {\cal C}|\Psi_{x}|^2/|\Psi_{x}^0|^2$ with ${\cal C}$ the normalization factor is approximated by using $\sum_x I_x / I^0 = 1$. This makes VMC sample efficient with DNN ansatzes.

For large systems, the number of samples from MCMC needs to be large for reliable numerical results. The memory of a single accelerator is the bottleneck.
Therefore, we scale up Eq.~(\ref{isgo}) on multiple accelerator workers as shown in Fig~\ref{fig:vmc_graph}(b). One cycle from sampler to workers and back is one {\it iteration}.
In each iteration, the MCMC sampler receives parameters $w$ in the current checkpoint and uses it to sample states and compute their corresponding $\ln\Psi$.
Duplicated states are grouped into unique states with counts.
The data are sharded and distributed back to the workers.
Workers are single-program multiple-data (SPMD) programs and each worker governs a set of accelerator devices.
Each worker computes local values $I_x$, $E_x$, $\partial_w\text{ln}\Psi_x$ in parallel. And global values $\sum_x I_x E_x$ and $\sum_x I_x$ are computed by the efficient all-reduce sum supported by many deep learning frameworks and hardware, e.g. \verb|psum| in JAX~\cite{jax2018github}.
Finally, the local gradients are also averaged over workers by the all-reduce mean and parameters on each work are updated identically. After a certain number of updating steps, a checkpoint is saved, and the sampler uses this new checkpoint to start the next iteration.

\section{Numerical Experiments}\label{results}
We demonstrate scalable VMC with GNA on hard-core Boson systems. The Hamiltonian is
\begin{equation}\label{eq:hamiltonian}
    H=-t\sum_{\langle ij \rangle}(b^{\dagger}_i b_j + b^{\dagger}_j b_i) + V\sum_{\langle ij \rangle}n_in_j,
\end{equation}
where $b^{\dagger}_i$ ($b_i$) are creation (annihilation) operators of hard-core Bosons on site $i$. Two hard-core Bosons can not occupy the same site. $n_i=b_i^{\dagger}b_i$ are Boson number operators on site $i$.
The first term in Eq.~(\ref{eq:hamiltonian}) is the tunneling, and the second term is the nearest neighbor interaction. The tunneling parameter $t > 0$ varies and the interaction strength $V=1$ is fixed. Since the off-diagonal elements in Eq.~(\ref{eq:hamiltonian}) are always negative, it can be proved that the ground state wavefunction is always positive semi-definite. So in this paper, we use GNAs with real-valued parameters only. 
To show the representation power of GNA, we consider the Hamiltonian on three different geometries with interesting physics properties, differing by the nearest neighbors $\langle ij \rangle$.
{\it Kagome lattice:} as $t/V$ increases, the system undergoes a valence bond solid to superfluid phase transition. Whether the phase transition is weakly first-order or a continuous superfluid-solid transition is still under investigation~\cite{Isakov:2006aa, Zhang:2018aa}.
{\it Triangular lattice:} as $t/V$ increases, the system undergoes a first-order supersolid to superfluid transition~\cite{Heidarian:2005aa, Melko:2005aa, Wessel:2005aa, Boninsegni:2005aa}. 
{\it randomly connected graph:} recently, disordered models such as Sachdev-Ye-Kitaev model~\cite{Sachdev:1993aa, Kitaev2015} received much attention. We consider a simple model of hard-core Bosons on a randomly connected graph. The probability of arbitrary two sites being connected is $0.5$. These geometries are shown in Fig~\ref{fig:lattices}. Throughout the experiments, we use 5 Markov chains each with length $10^4$, 500 sampling iterations, and 100 parameter updating steps within each sampling iteration. In practice, we find the energy often stops changing too much after 100 updating steps within one iteration. And the energy converges after a few sampling iterations (on the order of 10). Using more parameter updating steps or fewer samples per iteration may lead to gradient explosion, because if the initial sampling distribution is too far from the instantaneous wavefunction, reweighting will become catastrophically bad. However, the exploration of such hyperparameters is not the focus of this paper. \citet{yang2020deep} presents a comparison of the ISGO and conventional VMC algorithm on computational speed.

\subsection{Benchmarking ground state energies with exact diagonalization}
\begin{figure*}[htb!]
	\includegraphics[width=0.95\textwidth]{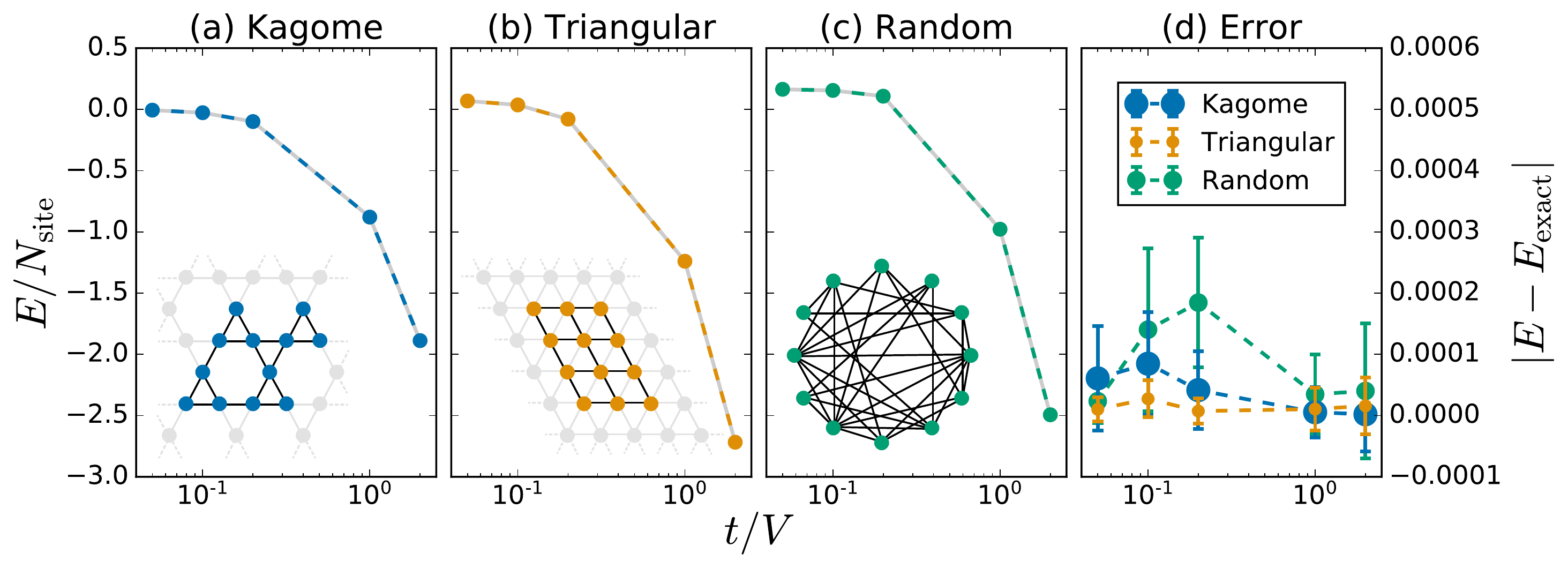}
	\centering
	\caption{Comparison of ground state energies between the VMC with GNA (color) and exact diagonalization (gray) for (a) the $2 \times 2 \times 3$ Kagome lattice, (b) the $3 \times 4$ triangular lattice and (c) randomly connected graph with $12$ sites.
	For Kagome and triangular lattices, the periodic boundary condition is enforced along with two primitive directions, so the lattice form a torus.}
	\label{fig:lattices}
\end{figure*}

In Fig~\ref{fig:lattices}, we compute the ground state energies using VMC with GNA for (a) a 2D Kagome lattice, (b) a triangular lattice, (c) randomly connected graphs. The mean and standard deviation of the ground state energy is calculated from the variational energies of the last 50 sampling iterations.
Since they are small systems, the exact results can be computed via exact diagonalization.
As shown in Fig~\ref{fig:lattices}(d), VMC with GNA consistently produces accurate ground-state energies comparing to the exact solutions among different geometries and $t/V$.

\begin{figure*}[htb!]
	\includegraphics[width=0.95\textwidth]{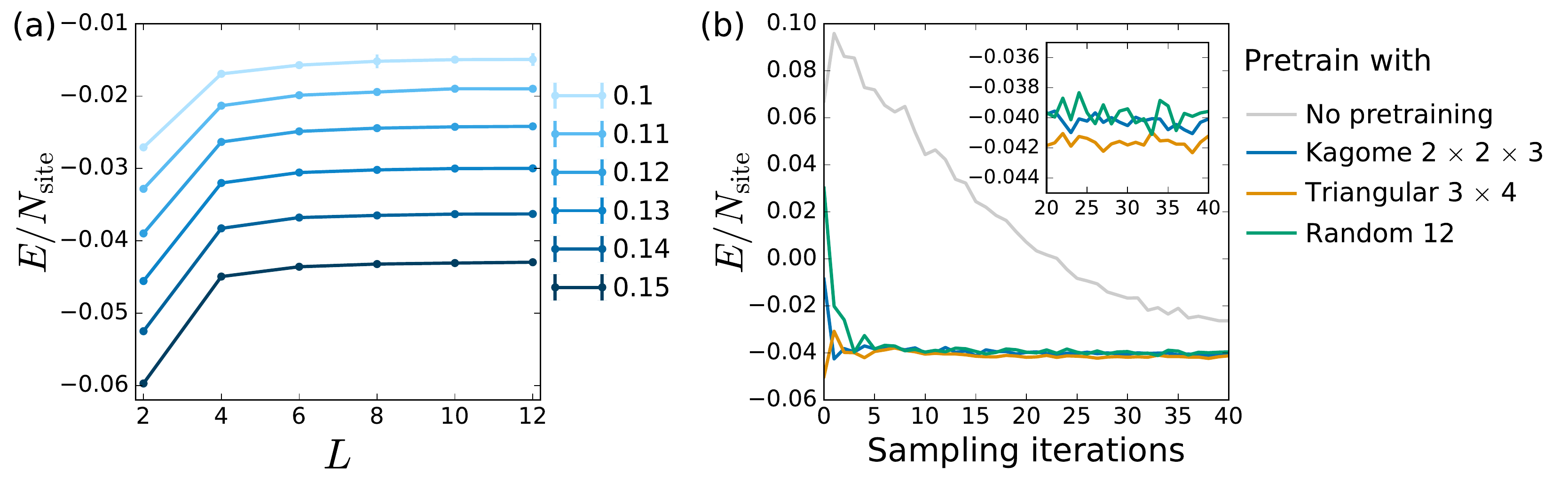}
	\centering
	\caption{(a) Ground state energies of Kagome lattice verses system sizes $L$. The total number of sites is $L \times L \times 3$. Different curves denote  various $t/V$. We use 8 TPU cores for $L < 8$ and up to 128 TPU cores for $L$ up to 12. (b) $12\times12\times3$ Kagome lattice with different pretraining from 12 sites.}
	\label{fig:system_sizes}
\end{figure*}

\subsection{Solving Kagome lattice with large system sizes}
Fig~\ref{fig:system_sizes}(a) shows the scalability of our approach, we compute the 2D Kagome lattice up to $12 \times 12 \times 3$, which is 432 sites in total. In each iteration, $5\times 10^4$ samples are generated. Since the memory of a typical GPU is about 16G, the previous ISGO method fails on this system as the sampled states and their intermediate activations in the GNA can not fit into a single device.

Another interesting property of GNA is that the trained parameters are transferable to different system sizes or even different geometries. Note that the second point is hard for CNN. 
It allows us to pretrain a GNA on small systems and apply it to larger systems or even other geometries. In Fig~\ref{fig:system_sizes}(b) we show that GNA pretrained on small systems with 12 sites significantly improves the convergence on the large system with 432 sites. Without pretraining, we observe that the GNA with randomly initialized parameters often leads to more unique states, or sometimes leads to gradient explosion.
With the GNA pretrained on a Kagome lattice with only $1/36$ the size of the targeted large Kagome lattice, the computation converges within 5 sampling iterations while GNA without pretraining does not even at 40. Interestingly, this effect stands even when the pretraining geometries, e.g. triangular and randomly connected graph, are different from the target geometries.

\section{Conclusion and future work}
We have combined GCN, a specific form of GNN, with a scalable implementation of the ISGO algorithm for calculating the ground state energies of hard-core Bosons on various geometries. 
As a universal graph encoder, GCN allows us to extend the previous neural network ansatz to systems with arbitrary geometry. Moreover, the parameter sharing nature of GCN enables fine-tuning of the model on a large system from the model pretrained on small systems.
With a scalable ISGO implementation, we compute ground state energies of Kagome lattices with size up to $12 \times 12 \times 3=432$  sites on 128 TPU cores. Given a trained GCN, we expect other physical quantities, such as correlation functions, can be computed for much larger system sizes. This will be helpful for exploring the quantum phase transition properties of the system.
In this paper, we study positive semi-definite ground state wavefunctions, thus GCN with real-valued parameters is enough. Ongoing work includes extending the method for complex-valued wavefunctions to resolve the wavefunction sign problem, which will be an unified framework that can be used for computing ground states of strongly interacting Fermions, frustrated spins, or computing unitary dynamics. This could be achieved by various approaches, for example, complex neural networks~\cite{PhysRevB.100.125124, Zimmermann2011ComparisonOT, 6789419, trabelsi2018deep} or two-head networks outputting the absolute value and angle.

\section*{Acknowledgement}
The authors thank Hanjun Dai and Bo Dai for helpful discussion. W.H. is supported by the U.S. Department of Energy (DOE), Office of Science, Basic Energy Sciences (BES), Materials Sciences and Engineering Division.

\section*{Broader Impact}
This research develops a scalable variational Monte Carlo (VMC) algorithm and a universal graph neural ansatz (GNA). The combination of them is applied to a fundamental scientific field --- quantum many-body physics.
Most of the ethical aspects and societal consequences are not applicable due to the fundamentality of this research.
However, we would like to address one possible negative outcome if this method received much attention and further developed.
Research organizations with rich computation resources, especially machine learning hardware accelerators, would benefit most from the idea proposed by this work.
By contrast, small organizations and individual researchers would have less or even no advantage to taking part in the development of this field. However, there may be other alternatives available, e.g. more affordable commercial cloud services for research or national supercomputer centers.

\bibliography{main}
\bibliographystyle{unsrtnat}
\end{document}